\documentclass{elsart}

\usepackage{natbib}

   \usepackage{graphicx}
   \usepackage{epstopdf}
  \usepackage{amssymb}

\def\teq#1{$\, #1\,$}                         
\def\aa{{Astron. Astrophys.}}

\def\apj{ApJ}
\def\apjsupp{ApJ Supp.}

\def\app{Astroparticle Phys.}                   
\def\apss{Astr. Space Sci.}                     

\def\mnras{{M.N.R.A.S.}}
\def\prl{Phys. Rev. Lett.}                      
\def\ssr{Space Sci. Rev.}                       
\def\reference{\par \noindent \hangafter=1 \hangindent=0.7 true cm}
             \font\sevenrm=cmr7

\def\dover#1#2{\hbox{${{\displaystyle#1 \vphantom{(} }\over{
   \displaystyle #2 \vphantom{(} }}$}}

\begin{document}
\newcommand{\vol}[2]{$\,$\rm #1\rm , #2,}                 
\def\gamsk{\gamma_1}
\def\erg{\varepsilon_\gamma}
\def\dover#1#2{\hbox{${{\displaystyle#1 \vphantom{(} }\over{
   \displaystyle #2 \vphantom{(} }}$}}
\def\thetascatt{\theta_{\hbox{\sevenrm scatt}}}
\def\thetaBone{\Theta_{\hbox{\sevenrm Bn1}}}
\def\thetaBtwo{\Theta_{\hbox{\sevenrm Bn2}}}
{\catcode`\@=11 
  \gdef\SchlangeUnter#1#2{\lower2pt\vbox{\baselineskip 0pt\lineskip0pt 
  \ialign{$\m@th#1\hfil##\hfil$\crcr#2\crcr\sim\crcr}}}} 
\def\gtrsim{\mathrel{\mathpalette\SchlangeUnter>}} 
\def\lesssim{\mathrel{\mathpalette\SchlangeUnter<}} 
\newcommand{\figureoutpdf}[5]{\centerline{}
   \centerline{\hspace{#3in} \includegraphics[width=#2truein]{#1}}
   \vspace{#4truein} \caption{#5} \centerline{} }
\begin{frontmatter}



\title{Gamma-Ray Burst Prompt Emission: Implications from Shock Acceleration Theory}

\author{Matthew G. Baring}

\address{Department of Physics and Astronomy, MS-108,
Rice University, P. O. Box 1892, Houston, TX 77251-1892, USA\\
{\tt Email: baring@rice.edu}}

\begin{abstract}
The principal paradigm for gamma-ray bursts suggest that the prompt
transient gamma-ray signal arises from multiple shocks internal to the
relativistic expansion.  This paper illustrates some properties of
diffusive acceleration at relativistic shocks that pertain to GRB
models, providing interpretation of the BATSE/EGRET data. Using a
standard Monte Carlo simulation, computations of the spectral shape, and
the range of spectral indices are presented, as functions of the shock
speed, magnetic field obliquity and the type of particle scattering.  It
is clear that while quasi-parallel, relativistic shocks with fields
approximately normal to the shock plane can efficiently accelerate
particles, highly oblique and perpendicular ones cannot unless the
particle diffusion is almost isotropic, i.e. extremely close to the Bohm
limit.  Accordingly, an array of distribution indices should be present
in the burst population, as is exhibited by the EGRET data, and even
mildly-relativistic internal shocks require strong field turbulence in
order to model the \teq{>100}MeV observations. In addition, recent
spectral fitting to burst data in the BATSE band is discussed, providing
probes of the efficiency of injection of non-thermal electrons.
\end{abstract}


\end{frontmatter}
\setlength{\parindent}{.25in}

\section{Introduction}
\label{sec:Introduction}

Gamma-ray bursts are among the most interesting and exotic phenomena in
astrophysics. In standard gamma-ray burst (GRB) models, the rapidly
expanding fireball cools, converting the internal energy of the hot
plasma into kinetic energy of the beamed, relativistically moving ejecta
and electron-positron pairs. At the point where the fireball becomes
optically thin and the GRB we see is emitted, the matter is too cool to
emit gamma-rays unless some mechanism can efficiently re-convert the
kinetic energy back into internal energy, i.e., unless some particle
acceleration process takes place. Diffusive shock (Fermi) acceleration
is widely believed to be this mechanism (e.g., Rees \& M\'esz\'aros
1992; Piran 1999). The large energy release in radiation from GRBs,
coupled to the limits on available energy from likely sources such as
supernovae or coalescing compact objects (see Piran 1999 and
M\'esz\'aros 2002 for reviews), requires that this acceleration of
particles at shocks be efficient.  Moreover, the rapid variability of
the prompt emission together with the impressive power of these sources
strongly suggests that their environs are moving ultrarelativistically
(e.g. Paczy\'nski 1986), an indication reinforced
by inferences from \teq{\gamma\gamma} transparency arguments (e.g.
Baring \& Harding 1997).  Therefore, a detailed understanding of 
particle acceleration at relativistic shocks is clearly motivated for 
GRB studies.

Acceleration of particles at non-relativistic shocks has been
extensively investigated in the contexts of supernova remnant and
heliospheric shocks. Diffusive acceleration at relativistic shocks is
less exhaustively studied than that for non-relativistic flows, yet it
may occur in extreme objects such as pulsar winds, hot spots in radio
galaxies, jets in active galactic nuclei and microquasars, and GRBs.
Early work on relativistic shocks was mostly analytical in the
test-particle approximation (e.g., Peacock 1981; Kirk \& Schneider
1987a; Heavens \& Drury 1988; Kirk \& Webb 1988). Some analytic work
(Schneider \& Kirk 1987; Baring \& Kirk 1991) has explored nonlinear,
cosmic ray modified shocks. Complementary Monte Carlo techniques have
been employed for relativistic shocks by a number of authors, including
test-particle analyses by Kirk \& Schneider 1987b and Ellison, Jones \&
Reynolds (1990) for parallel, steady-state shocks, and extensions to
include oblique magnetic fields by Ostrowski (1991), Ballard \& Heavens
(1992) and Bednarz \& Ostrowski (1998).

A key characteristic that distinguishes relativistic shocks from their
non-relativistic counterparts is their inherent anisotropy due to rapid
convection of particles through and away downstream of the shock, since
particle speeds \teq{v} are never much greater than the downstream flow
speed \teq{u_2\sim c/3}. Accordingly, the diffusion approximation, the
starting point for virtually all analytic descriptions of acceleration
at non-relativistic shocks, cannot be invoked since it requires nearly
isotropic distribution functions.  Hence analytic approaches prove more
difficult for relativistic shocks, though advances in special cases such
as the limit of extremely small angle scattering ({\it pitch angle
diffusion}) are possible (Kirk \& Schneider 1987a; Kirk, et al. 2000;
Keshet \& Waxman 2004). This paper explores some of the distinctive
properties of particle acceleration at relativistic shocks that are
germane to the gamma-ray burst paradigm, with a focus on spectral
issues, specifically the slope of the power-law distribution and the
efficiency of generation of this non-thermal component.

\section{Acceleration at Relativistic Shocks in GRBs}
 \label{sec:relshock}

A most attractive feature of non-relativistic shock acceleration
theory is that the distribution of accelerated particles is
scale-independent, i.e. a power-law.  This is a consequence of high
energy particles (those with speeds \teq{v \gg u_1}, with \teq{u_1}
being the upstream flow speed) attaining isotropy in all pertinent
reference frames.  At such energies, the principal transport equation
describing the acceleration process, the diffusion-convection equation,
can be solved analytically for plane shocks (e.g., Blandford \& Ostriker
1978; Jones \& Ellison 1991), yielding the well-known result for the
momentum distribution 
\begin{equation}
   \dover{dn}{dp} \, \propto\, p^{-\sigma} 
   \quad \hbox{with} \quad
   \sigma = \dover{r+2}{r-1} \ ,
 \label{eq:TPpowerlaw}
\end{equation}
where \teq{r=u_1/u_2} is the shock (velocity) compression ratio, \teq{p}
is the momentum.  Eq.~(\ref{eq:TPpowerlaw}) is a steady-state,
test-particle result, and the index \teq{\sigma} {\it depends only on
the velocity compression ratio} \teq{r}, i.e. hydrodynamic quantities. 
This elegant result does not carry over to relativistic shocks because
of their strong plasma anisotropy.  As a consequence, while power-laws
are in fact created, the index \teq{\sigma} becomes a function of the
flow speed, the field obliquity, and the nature of the scattering, all
of which intimately control the degree of particle anisotropy.

In the specific case of parallel (i.e. those where the field is locally
normal to the shock surface), ultrarelativistic shocks, the analytic
work of Kirk et al. (2000) demonstrated that as
\teq{\Gamma_1=1/\sqrt{1-(u_1/c)^2}\,\to\infty}, the spectral index
\teq{\sigma} asymptotically approached a constant, \teq{\sigma\to 2.23}
(see also Keshet \& Waxman 2004), a value realized once
\teq{\Gamma_1\gtrsim 10}.  This captivating result has been confirmed by
Monte Carlo simulations (Bednarz \& Ostrowski 1998; Baring 1999,
Achterberg et al. 2001; Ellison \& Double 2002), and corresponds to the
special case of compression ratios of \teq{r=3} and the particular
assumption of small scattering (pitch angle diffusion), specifically for
incremental changes \teq{\thetascatt} in a particle's momentum with
angle \teq{\thetascatt \ll 1/\Gamma_1}.  Here we explore departures from
this particular case that are appropriate for burst studies.

\subsection{Relativistic Shocks: an Array of Spectral Indices}
 \label{sec:index_array}

The spectral index of the power-law distribution is a declining function
of the Lorentz factor \teq{\Gamma_1} for a fixed compression ratio, a
characteristic evident in Kirk \& Schneider (1987a), Ballard \& Heavens
(1991), and Kirk et al. (2000) for the case of pitch angle scattering,
and a property that extends to large angle scattering (Ellison, Jones \&
Reynolds 1990; Baring, 1999; Ellison \& Double 2004). Faster shocks, if
parallel, generate flatter distributions if \teq{r} is held constant
(e.g. see the parallel shock, \teq{\thetaBone =0^\circ} case in
Fig.~\ref{fig:obq_index}), a consequence of the increased kinematic
energization occurring at relativistic shocks.  A tabulation of this
property, namely \teq{\sigma} values for different \teq{r} in the case
of pitch angle diffusion, is given in Baring (2004), and analytic
approximations are derived in Keshet \& Waxman (2004). Note that
imposing a specific equation of state such as the J\"uttner-Synge one
renders \teq{r} a function of \teq{\Gamma_1} so that this monotonicity
property can disappear, as evinced in Fig.~2 of Kirk et al. (2000; see
also Keshet \& Waxman 2004).  Moreover, for significant field
obliquities, the trend with shock speed is reversed, as is evident in
Fig.~\ref{fig:obq_index}.

Astrophysical models usually invoke the canonical compression ratio
\teq{r=3}, the well-known result for a relativistic, purely hydrodynamic
shock possessing an ultrarelativistic equation of state.  However, in
cases where the magnetic field becomes dynamically important (e.g. the
termination shock for the Crab pulsar wind: Kennel \& Coroniti 1984),
the strong fields can weaken magnetohydrodynamic shocks considerably,
just as they do in non-relativistic situations. Moreover, in
ultrarelativistic shocks, where pressure anisotropy can be significant,
Double et al. (2004) observed that the shock could be strengthened {\it
or} weakened, depending on the nature of the pressure anisotropy, which
must be a significant function of the angle, \teq{\thetaBone}, the
upstream field makes to the shock normal. Hence, it is anticipated that
\teq{\sigma} will be a function of \teq{\thetaBone}.  The influence of
pressure anisotropy on the shock compression will be greatest in cases
where the index is low, i.e. \teq{\sigma\lesssim 2}, so that the
accelerated particles can be influential in determining the dynamics of
the (non-linear) shock.

Also of interest is the fact that the slope of the nonthermal particle
distribution depends on the nature of the scattering, a feature evident
in the works of Ellison, Jones \& Reynolds (1990), Bednarz \& Ostrowski
(1998) and Baring (1999) .  The asymptotic, ultrarelativistic index of
2.23 is realized only in the mathematical limit of pitch angle diffusion
(PAD), where the particle momentum is stochastically deflected on
arbitrarily small angular (and therefore temporal) scales.  In practice,
PAD results when the scattering angle \teq{\thetascatt} is smaller than
the Lorentz cone angle \teq{1/\Gamma_1} in the upstream region.  In such
cases, particles diffuse in the region upstream of the shock only until
their angle to the shock normal exceeds around  \teq{1/\Gamma_1}.  Then
they are rapidly swept to the downstream side of the shock.  

\begin{figure}[htb]
\figureoutpdf{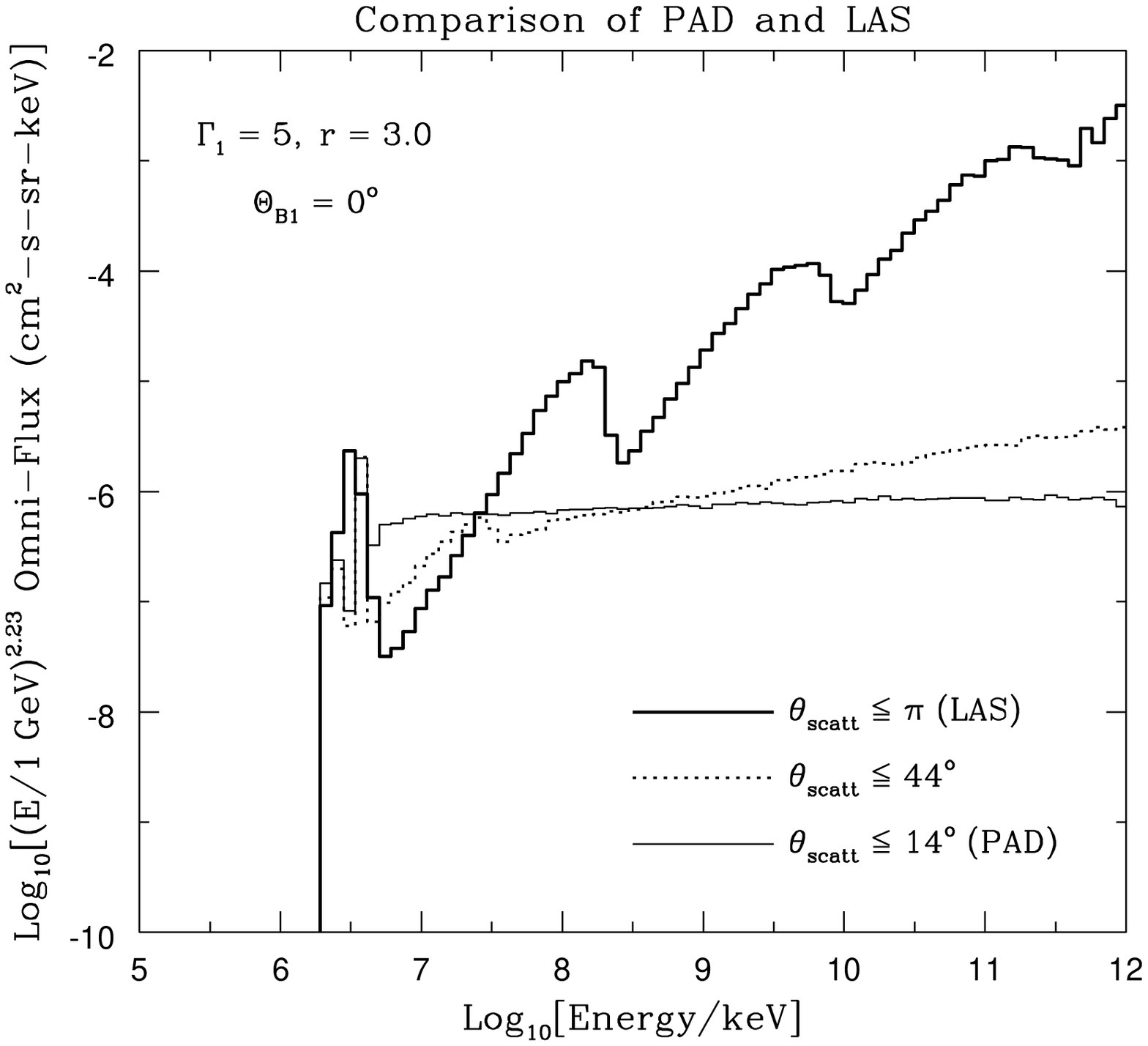}{5.5}{0.0}{-0.75}{
Particle distributions from a parallel (\teq{\thetaBone=0}) relativistic
shock of \teq{r=3} and Lorentz factor \teq{\Gamma_1=5}, obtained from a
Monte Carlo simulation (Ellison, Jones \& Reynolds 1990; Baring, 1999). 
Scattering is modeled by randomly deflecting particle momenta by an
angle \teq{\thetascatt} within a cone whose axis coincides with the
momentum prior to scattering.  Distributions are depicted for three
cases, \teq{\thetascatt \leq 14^\circ}, corresponding to pitch angle
diffusion (PAD), large angle scattering (LAS: \teq{\thetascatt\leq\pi\gg
1/\Gamma_1}), and an intermediate case (dotted histogram). The
distributions are divided by the \teq{E^{-2.23}} ultra-relativistic
power-law determined by Kirk et al. (2000).
}
 \label{fig:spectra}
\end{figure}

Particle distributions obtained from the Monte Carlo simulation of
acceleration at relativistic shocks developed by Ellison, Jones \&
Reynolds (1990) are exhibited in Fig.~\ref{fig:spectra}.  They
demonstrated that for large angle scattering (LAS, with
\teq{\thetascatt\sim\pi}) the spectrum is highly structured and much
flatter than \teq{E^{-2}}.  Such a case is exhibited in the Figure.  The
structure is kinematic in origin, where large angle deflections lead to
distribution of fractional energy gains between unity and
\teq{\Gamma_1^2}.  Gains like this are kinematically analogous to the
energization of photons by relativistic electrons in inverse Compton
scattering, and are much larger on average than those realized in PAD
(see Gallant \& Achterberg 1999; Baring 1999). Each structured spectral
segment in Fig.~\ref{fig:spectra} corresponds to an increment in the
number of shock crossings, successively from \teq{1\to3\to 5\to 7} etc.,
as illustrated by Baring (1999), that eventually smooth out to
asympotically approach an index of \teq{\sigma\sim 1.5}. Clearly, such
highly-structured distributions have not been inferred from radiation
emission in gamma-ray bursts or any other astrophysical objects.  Note
that the \teq{\Gamma_1=5} results depicted here are entirely
representative of ultrarelativistic shocks.

An intermediate case is also depicted in Fig.~\ref{fig:spectra}, with
\teq{\thetascatt\sim 4/\Gamma_1}.  The spectrum is smooth, like the PAD
case, but the index is lower than 2.23. Magnetic turbulence could easily
be sufficient to effect scatterings on the order of these angles, a
contention that becomes even more salient for ultrarelativistic shocks
with \teq{\Gamma_1\gg 10}.  Clearly a range of indices can be supported
when \teq{\thetascatt} is chosen to be of the order of \teq{1/\Gamma_1},
and the scattering corresponds to the transition between the PAD and LAS
limits. Hence, it is expected that various astrophysical systems will
encompass a range of scattering properties.  Accordingly, the continuous
and monotonically decreasing behavior of \teq{\sigma} with
\teq{\thetascatt}, as indicated in the exposition of Ellison \& Double
(2004), highlights the significant range of distribution indices in
relativistic shocks.

In relativistic shocks, the diversity of spectral indices is enhanced by
transport properties orthogonal to the mean field direction. The
diffusion of particles across mean field lines, becomes a critical
element in the discussion of oblique or perpendicular shocks.  In
non-relativistic shocks, when the upstream angle \teq{\thetaBone} of the
field to the shock normal is significant,  diffusion of particles in the
downstream region struggle to compete with convective losses, and
transport back upstream of the shock layer becomes inefficient.   This
leads to the quenching of injection of thermal particles of speed
\teq{v\gtrsim u_1}, which fail to return to the shock after one crossing
to the downstream side.  Accordingly, for \teq{\thetaBone\gtrsim
30^\circ}, the Fermi acceleration process ceases when transport across
field lines is suppressed (Baring, Ellison \& Jones 1994), and for
\teq{\thetaBone < 30^\circ}, while power-law superthermal distributions
are realized, their normalization is a strongly declining function of
\teq{\thetaBone}.

In relativistic shocks, this phenomenon translates to a dramatic
steepening of the distribution above thermal energies. When \teq{u_1\sim
c}, an oblique shock is inherently superluminal, so that convective
losses are pervasive for {\it all} particle speeds, not just slightly
suprathermal ones.  Such losses must diminish the nonthermal population,
and since the loss rate is purely a function of particle speed \teq{v}
(Peacock 1981; Jones \& Ellison 1991), which is effectively pinned at
\teq{v\approx c},  and \teq{u_2}, the overall effect is to increase the
spectral index while retaining power-law character.  Monte Carlo model
indices are illustrated in Fig.~\ref{fig:obq_index}, where the
simulation output was acquired in the absence of cross field diffusion
(i.e. for perpendicular spatial diffusion coefficient
\teq{\kappa_{\perp}=\lambda_{\perp}v/3 =0}, corresponding to weak
turbulence). Increasing \teq{\thetaBone} results in a rapid rise in
\teq{\sigma} corresponding to a suppression of acceleration.
Essentially, for \teq{\thetaBone\gtrsim 25^\circ}, acceleration is
virtually non-existent for \teq{\Gamma_1u_1/c\gtrsim 1}.  Note that
\teq{\kappa_{\parallel} =\lambda_{\parallel} v/3} is the component of
the spatial diffusion coefficient parallel to the field, so that
\teq{\kappa_{\parallel}/u_1} defines the effective spatial scale of
diffusion along the field lines.

\begin{figure}[htb]
\figureoutpdf{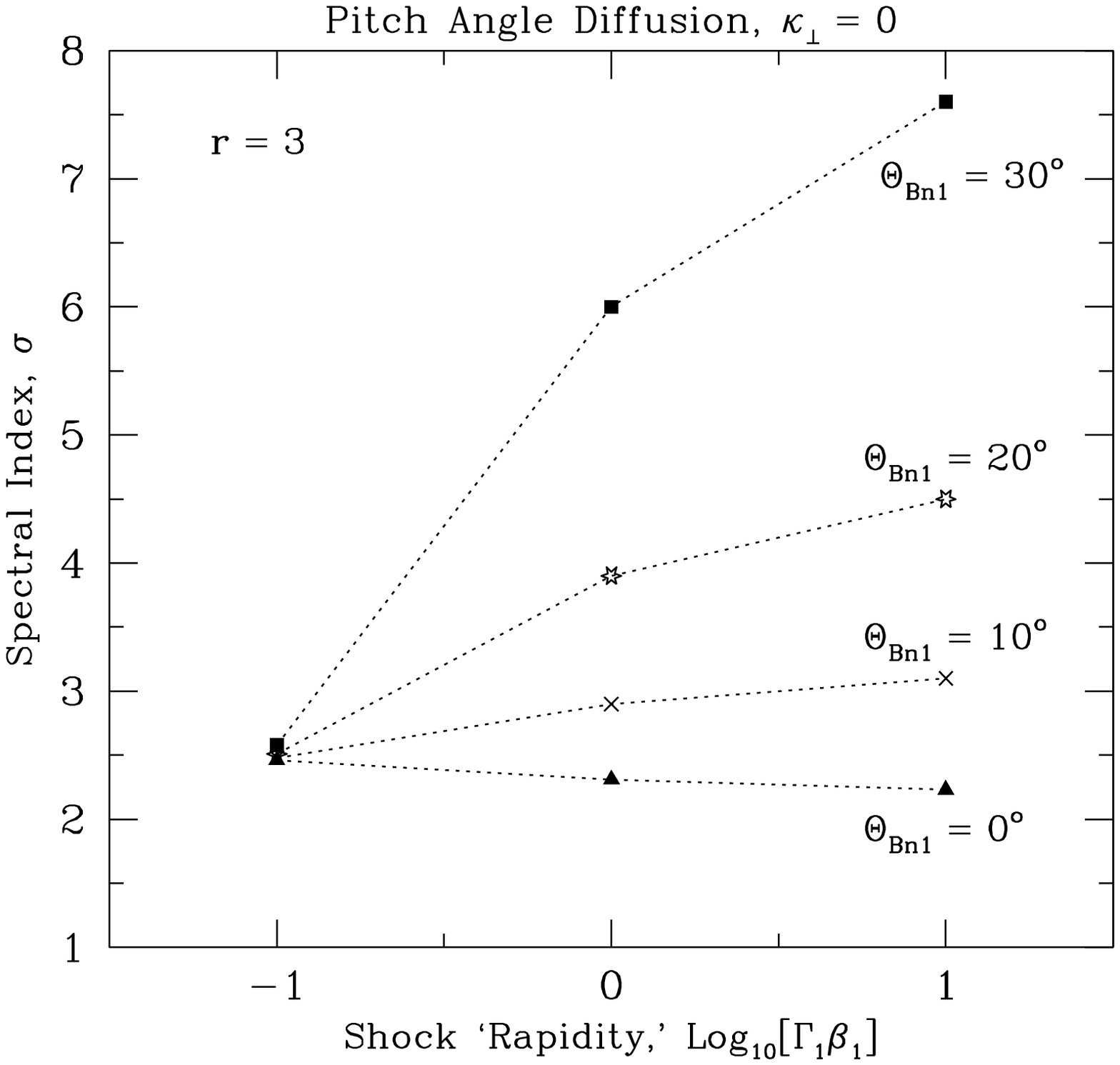}{5.5}{0.0}{-0.75}{
Particle distribution indices \teq{\sigma}, for \teq{dn/dp\propto p^{-\sigma}}, 
from oblique (\teq{\thetaBone >0}) relativistic shocks of \teq{r=3} as 
a function of \teq{\Gamma_1\beta_1}, obtained from a Monte Carlo 
simulation (Ellison, Jones \& Reynolds 1990; Baring 1999) in the 
limit of pitch angle diffusion.
Results are depicted for the case of zero diffusive transport
perpendicular to the mean field, i.e., \teq{\kappa_{\perp}=0} for the
component, perpendicular to {\bf B}, of the spatial diffusion
coefficient \teq{\kappa = \lambda v/3}, where \teq{\lambda} is a
particle's diffusive mean free path.  The index is insensitive to
\teq{\thetaBone} for non-relativistic shocks, but rapidly increases with
obliquity for relativistic ones, underlining their inherent inefficiency.
}
 \label{fig:obq_index} 
\end{figure}

If the shocked plasma is strongly turbulent, the system will be driven
towards the {\it Bohm-diffusion} limit, where diffusion coefficients are
similar parallel and perpendicular to the field, i.e.
\teq{\kappa_{\perp}\sim \kappa_{\parallel}}, and transport is
effectively isotropic.  Efficient transport across field lines returns
the particles to the shock from the downstream region, effectively
circumventing convective losses, and accordingly flattening the
power-law distribution.  In such cases, the strength of the turbulence
renders quasi-perpendicular shocks much like parallel ones, and the
distinction of field obliquity becomes less meaningful.  It is
anticipated that transport near the Bohm limit would be essential to
generate \teq{\sigma\lesssim 3}, i.e. indices relevant for acceleration
applications to burst models.  This is borne out in Bednarz \& Ostrowski
(1998) and the recent work of Ellison \& Double (2004), whose
test-particle Monte Carlo simulation results of exhibited the expected
monotonic decrease in \teq{\sigma} with an increase in
\teq{\kappa_{\perp}/ \kappa_{\parallel}}.

It must be remarked that the transport of particles in the shock
environment possesses a complex dependence on the nature of the dynamic
fields.  Niemiec \& Ostrowski (2004) very recently explored the
acceleration process at relativistic shocks by computing particle
trajectories near a shock in the presence of injected, strong field
turbulence.  They found that (i) the spectral index was dependent on the
strength of the turbulence, and more interestingly, (ii) that large
perturbation amplitudes \teq{\delta B/B\gtrsim 1} could actually render
a steepening of the distribution of accelerated particles for oblique,
subluminal shocks.  Niemiec \& Ostrowski (2004) contended that this
characteristic was largely a consequence of reduced reflection in the
shock layer, and observed that the effect was generally reversed in
oblique, superluminal shocks, where increasing \teq{\delta B/B}
flattened the distribution somewhat.  Their investigations, and the
Monte Carlo technique, of course do not explore quasi-coherent
electrodynamic acceleration that results from charge separation
potentials and currents in the shock layer; such plasma physics
properties are best probed using particle-in-cell (PIC) simulations,
which have been recently exploited by Silva et al. (2003) and Nishikawa
et al. (2003) to study the development of the Wiebel instability and
associated acceleration in relativistic plasma shocks.  These
injected turbulence and PIC 
simulations add to the evidence that the spectrum of particles
accelerated at relativistic shocks is not universal.

\subsection{Modeling Observed GRB Prompt Emission}
 \label{sec:model}

The sensitivity of the power-law index to shock obliquity and the nature
of turbulent transport immediately indicate that GRB spectra should
possess a diversity of indices, if a shock acceleration model is
applicable to the burst paradigm.  This is in fact manifested in the
energy range well above the \teq{\sim 1}MeV peak of emission in data
taken from the EGRET experiment on the Compton Gamma-Ray Observatory
(CGRO), where the half dozen or so bursts seen at high energies have a
broad range of spectral indices (Dingus 1995), namely \teq{\alpha \sim
2-3.7} for \teq{dn/d\erg\propto \erg^{-\alpha}}. This result suffers
from limited statistics due to (i) the nature of bursts, and (ii) to
EGRET's field of view being more constrained than that for BATSE, the
principal GRB experiment on CGRO.  The GLAST mission will provide a more
refined determination of the distribution of burst indices above 30 MeV
after its launch in 2007.

The EGRET data suggest that relatively flat indices are more common,
though there is an obvious observational bias against distinguishing
\teq{\alpha \gtrsim 6} cases, since poor statistics at the uppermost
energies will degrade index determination in such cases.  If particle
acceleration in bursts is indeed this efficient, then highly oblique
shocks cannot be present, unless the turbulence is extremely strong; the
results of Niemiec \& Ostrowski (2004) support this contention.  For
prompt EGRET emission, the prevailing scenario is that shocks internal
to the GRB flow/blast wave are responsible for the dissipative
conversion of bulk kinetic energy to observable radiation (e.g. see
Piran 1999; Meszaros 2002). These shocks are necessarily
mildly-relativistic, so \teq{\sigma \sim 3-4} values are reasonable for
near the Bohm diffusion, even if the shocks are highly oblique.  In
contrast, optical afterglow models invoke the outer blast wave shock as
the site for their energization, and in all probability, this
ultrarelativistic shock is quasi-perpendicular. In such a shock
environment, it is difficult to generate indices \teq{\sigma\sim 3} even
near the Bohm limit, so it is unclear how well \teq{\Gamma_1\gg 1}
shocks will be able to model burst afterglow spectra.

The power-law index is not the only acceleration characteristic germane
to the GRB problem: the shapes of the particle distributions at thermal
and slightly suprathermal energies are also pertinent.  This energy
domain samples particle injection or dissipational heating in the shock
layer, and is readily probed for electrons by the spectrum of prompt GRB
emission in the BATSE band.  Tavani (1996) obtained impressive spectral
fits to several bright BATSE bursts using a phenomenological electron
distribution and the synchrotron emission mechanism. This has been a
driver for the interpretation of burst spectra.  There are issues with
fitting low energy (i.e. \teq{\lesssim 100}keV) spectra in about 1/3 of
bursts (e.g. Preece et al. 1998) in the synchrotron model, yet this
emission mechanism still remains the most popular candidate today.
Tavani's work did not directly address theoretical characteristics of
distributions of accelerated particles at slightly suprathermal energies.

Perspectives based on acceleration theory underpinned the recent
analysis of Baring \& Braby (2004), who pursued a program of spectral
fitting of GRB emission using a linear combination of thermal and
non-thermal electron populations. These fits demanded that the
preponderance of electrons that are responsible for the prompt emission
reside in an intrinsically non-thermal population. This requirement
strongly contrasts particle distributions obtained from acceleration
simulations, exemplified by those depicted in Fig.~\ref{fig:spectra}.
The consequence is obviously a potential conflict for acceleration
scenarios where the non-thermal electrons are drawn directly from a
thermal gas (the virtually ubiquitous case), unless radiative
efficiencies only become significant at highly superthermal energies. 
This does not necessarily mean that acceleration at relativistic shocks
is not operating in bursts. It is possible that somehow, relativistic
shocks can suppress thermalization of electrons, though such a
conjecture has no definitive simulational evidence to support it at
present. A potential resolution to this dilemma is that strong radiative
self-absorption could be acting, in which case the BATSE spectral probe
is not actually sampling the thermal electrons.  It is also possible
that other radiation mechanisms such as pitch-angle synchrotron, or
jitter radiation may prove more desirable.  A goal of future work will
be to ascertain whether a shock acceleration paradigm can be truly
consistent with the GRB emission that is observed.

\section{Conclusions}
 \label{sec:conclusion}

This paper has discussed two key acceleration issues for models of
gamma-ray burst sources, namely expectations for the power-law index
\teq{\sigma} of the non-thermal population, and the efficiency for which
these particles are injected into the acceleration process. The recent
analysis of Kirk et al. (2000), applicable for the specific case of
pitch angle diffusion at plane parallel shocks, has demonstrated that
the index of this power-law asymptotically approaches \teq{\sigma =2.23}
as the Lorentz factor \teq{\Gamma_1} of the upstream flow in the shock
rest frame tends to infinity. Here it is illustrated how this
widely-quoted result is not universal, and that an array of indices are
possible in general.  Results from a kinetic Monte Carlo simulation of
diffusive shock acceleration are presented. These indicate that the
value of \teq{\sigma} is sensitive to the nature of the scattering,
becoming much smaller than \teq{2.23} when the particle transport
experiences angular deflections larger than around \teq{1/\Gamma_1}. 
The simulated distributions in \teq{\Gamma_1\gg 1} cases exhibit a rapid
steepening (i.e., higher \teq{\sigma}) as the magnetic field obliquity
\teq{\thetaBone} of the shock rises above zero, in the particular case
where diffusion perpendicular to the mean field is absent: efficient
acceleration is effectively quenched in moderately oblique and
quasi-perpendicular truly relativistic shocks unless there is strong
cross field diffusion.  The array of possible values of \teq{\sigma} is
commensurate with EGRET data on a handful of bursts.  This
characteristic favors an internal shock model for the prompt emission,
where the shocks are only mildly-relativistic in the comoving expansion
frame, and the spread of \teq{\sigma} values is narrower.  It is also
noted that the acceleration simulations only inject a minority of
thermal particles into the diffusive acceleration process, contrasting
electron distributions inferred in spectral fits for bright CGRO bursts.
This poses a potential conflict for acceleration models that motivates
further exploration of the energization of electrons via detailed
hydrogenic and pair plasma simulations of shock acceleration.

\section{References}


\setlength{\parskip}{.00in}

\reference
Achterberg, A., Gallant,ÊY.~A.; Kirk,ÊJ.~G., Guthmann, A.~W. Particle acceleration by 
   ultrarelativistic shocks: theory and simulations. \mnras\vol{328}{393--408} 2001.
\reference
Ballard, K.~R., Heavens, A.~F. First-order Fermi acceleration at oblique 
   relativistic magnetohydrodynamic shocks. \mnras\vol{251}{438--448} 1991.
\reference
Ballard, K.~R.,, Heavens, A.~F. Shock acceleration and steep-spectrum 
   synchrotron sources. \mnras\vol{259}{89--94} 1992.
\reference
Baring, M. G. Acceleration at Relativistic Shocks in Gamma-Ray Bursts.
   in \it Proc. of the 26th ICRC, Vol. IV \rm ,  p.~5--8, {\tt [astro-ph/9910128]}, 1999.
\reference
Baring, M. G.  Nucl. Phys. B Proc. Supp. \vol{136}{198--207}
   Diffusive Shock Acceleration of High Energy Cosmic Rays.  {\tt [astro-ph/0409303]}, 2004.
\reference
Baring, M.~G., Braby, M.~L. A Study of Prompt Emission Mechanisms 
   in Gamma-Ray Bursts. \apj\vol{613}{460--476} 2004.
\reference
Baring, M.~G., Ellison, D.~C., Jones, F.~C. Monte Carlo Simulations of
   Particle Acceleration at Oblique Shocks. \apjsupp\vol{90}{547--552} 1994.
\reference
Baring, M.~G., Harding, A.~K. The Escape of High-Energy Photons from 
   Gamma-Ray Bursts. \apj\vol{491}{663--680} 1997.
\reference
Baring, M.~G., Kirk, J.~G. The modification of relativistic shock fronts 
   by accelerated particles. \aa\vol{241}{329--342} 1991.
\reference
Bednarz, J., Ostrowski, M. Energy Spectra of Cosmic Rays Accelerated at 
   Ultrarelativistic Shock Waves. \prl\vol{80}{3911--3914} 1998.
\reference
Blandford, R.~D., Ostriker, J.~P. Particle acceleration by astrophysical shocks.
   \apj\vol{221}{L29--L32} 1978.
\reference
Dingus, B.~L. EGRET Observations of  $>30$ MeV Emission from the Brightest 
   Bursts Detected by BATSE. \apss\vol{231}{187--190} 1995.
\reference
Double, G.~P., Baring, M.~G., Jones, F.~C., Ellison, D.~C. 
   Magnetohydrodynamic Jump Conditions for Oblique Relativistic Shocks 
   with Gyrotropic Pressure.  \apj\vol{600}{485--500} 2004.
\reference
Ellison, D.~C., Double, G.~P. Nonlinear particle acceleration in relativistic 
   shocks. \app\vol{18}{213--228} 2002.
\reference
Ellison, D.~C., Double, G.~P. Diffusive shock acceleration in unmodified 
   relativistic, oblique shocks. \app\vol{22}{323--338} 2004.
\reference
Ellison, D.~C., Jones, F.~C., Reynolds, S.~P. First-order Fermi particle 
   acceleration by relativistic shocks. \apj\vol{360}{702--714} 1990.
\reference
Gallant, Y.~A., Achterberg, A. Ultra-high-energy cosmic ray acceleration 
   by relativistic blast waves. \mnras\vol{305}{L6--L10} 1999.
\reference
Heavens, A.~F., Drury, L.~O'C.  Relativistic shocks and particle acceleration.
   \mnras\vol{235}{997--1009} 1988.
\reference
Jones, F.~C., Ellison, D.~C. The plasma physics of shock acceleration. 
   \ssr\vol{58}{259--346} 1991.
\reference
Kennel, C.~F., Coroniti, F.~V. Confinement of the Crab pulsar's wind 
   by its supernova remnant. \apj\vol{283}{694--709} 1984.
\reference
Keshet, U., Waxman, E. The spectrum of particles accelerated in relativistic,
   collisionless shocks. preprint, {\tt [astro-ph/0408489]}, 2004.
\reference
Kirk, J.~G., Guthmann, A.~W., Gallant, Y.~A., Achterberg, A. Particle Acceleration 
   at Ultrarelativistic Shocks: An Eigenfunction Method. \apj\vol{542}{235--242} 2000.
\reference
Kirk, J.~G., Schneider, P. On the acceleration of charged particles 
   at relativistic shock fronts. \apj\vol{315}{425--433} 1987a.
\reference
Kirk, J.~G., Schneider, P. Particle acceleration at shocks - 
   A Monte Carlo method. \apj\vol{322}{256--265} 1987b.
\reference
Kirk, J.~G.,  Webb, G.~M. Cosmic-ray hydrodynamics at relativistic shocks.
   \apj\vol{331}{336--342} 1988.
\reference
M\'esz\'aros, P. Ann. Rev. Astron. Astr. Theories of Gamma-Ray Bursts.
   \vol{40}{137--169} 2002.
\reference
Niemiec, J., Ostrowski, M. Cosmic-Ray Acceleration at Relativistic Shock Waves 
   with a ``Realistic'' Magnetic Field Structure. \apj\vol{610}{851--867} 2004.
\reference
Nishikawa, K.-I., Hardee,ÊP., Richardson,ÊG., et al. Particle Acceleration in 
   Relativistic Jets Due to Weibel Instability. \apj\vol{595}{555--563} 2003.
\reference
Ostrowski, M. Monte Carlo simulations of energetic particle transport in 
   weakly inhomogeneous magnetic fields. I - Particle acceleration in relativistic 
   shock waves with oblique magnetic fields.  \mnras\vol{249}{551--559} 1991.
\reference
Paczy\'nski, B. Gamma-ray bursters at cosmological distances. 
   \apj\vol{308}{L43--L46} 1986.
\reference
Peacock, J.~A. Fermi acceleration by relativistic shock waves. 
   \mnras\vol{196}{135--152} 1981.
\reference
Piran, T.  Phys. Rep. Gamma-ray bursts and the fireball model.
   \vol{314}{575--667} 1999.
\reference
Preece, R.~D., Briggs,ÊM.~S., Mallozzi,ÊR.~S., et al. The Synchrotron 
   Shock Model Confronts a ``Line of Death'' 
   in the BATSE Gamma-Ray Burst Data. \apj\vol{506}{L23--L26} 1998.
\reference
Rees, M.~J., M\'esz\'aros, P. Relativistic fireballs -- Energy conversion 
   and time-scales. \mnras\vol{258}{41--43} 1992.
\reference
Schneider, P., Kirk, J. G. Fermi acceleration at shocks with arbitrary 
   velocity profiles. \apj\vol{323}{L87--L90} 1987.
\reference
Silva, L.~O., Fonseca,ÊR.~A., Tonge,ÊJ.~W., et al. Interpenetrating Plasma 
   Shells: Near-equipartition Magnetic Field Generation and Nonthermal 
   Particle Acceleration. \apj\vol{596}{L121--L124} 2003.
\reference
Tavani, M. Shock Emission Model for Gamma-Ray Bursts.
   \prl\vol{76}{3478--3481} 1996.

\end{document}